\documentclass[10pt,conference]{IEEEtran}
\usepackage{amssymb,color,amsmath} 
\usepackage{cite}
\usepackage{paralist}

\DeclareMathOperator{\tr}{tr}

\newcommand{\C}{\mathbf{C}}
\newcommand{\F}{\mathbf{F}}
\newcommand{\wt}{{\rm {wt}}}
\newcommand{\ord}{{\rm {ord}}}
\newcommand{\nix}[1]{}
\newcommand{\hdual}{{\bot_h}}
\newcommand{\qr}{q\equiv\square\bmod{n}}
\newcommand{\qrx}{q\equiv\square\bmod}
\newcommand{\ket}[1]{\left|{#1}\right\rangle}

\newtheorem{theorem}{Theorem}
\newtheorem{lemma}[theorem]{Lemma}

\newtheorem{example}[theorem]{Example}

\begin{document}

\title{Remarkable Degenerate Quantum Stabilizer Codes Derived from 
Duadic Codes}

\author{\authorblockN{Salah A.~Aly, Andreas Klappenecker, Pradeep
Kiran Sarvepalli}
\authorblockA{Department  of Computer Science,
Texas A\&M University, College Station, TX 77843-3112, USA \\
Email: \{salah,klappi,pradeep\}@cs.tamu.edu}}
\maketitle

\begin{abstract}
Good quantum codes, such as quantum MDS codes, are typically
nondegenerate, meaning that errors of small weight require active
error-correction, which is---paradoxically---itself prone to
errors. Decoherence free subspaces, on the other hand, do not require
active error correction, but perform poorly in terms of minimum
distance.  In this paper, examples of degenerate quantum codes are
constructed that have better minimum distance than decoherence free
subspaces and allow some errors of small weight that do not require
active error correction.  In particular, two new families of
$[[n,1,\geq \sqrt{n}]]_q$ degenerate quantum codes are derived from
classical duadic codes.
\end{abstract}

\section{Introduction}\label{sec:intro}
Suppose that $q$ is a power of a prime $p$. Recall that an
$[[n,k,d]]_q$ quantum stabilizer code $Q$ is a $q^k$-dimensional
subspace of ${\C^{q^n}}$ such that $\langle u|E|u\rangle=\langle
v|E|v\rangle$ holds for any error operator $E$ of weight $\wt(E)<d$
and all $\ket{u}, \ket{v} \in Q$, see~\cite{ashikhmin01,ketkar05} for
details.  The stabilizer code $Q$ is called nondegenerate (or pure) if
and only if $\langle v|E|v\rangle=q^{-n}\tr E$ holds for all errors
$E$ of weight $\wt(E)<d$; otherwise, $Q$ is called degenerate. Recall
that purity and nondegeneracy are equivalent notions in the case of
stabilizer codes, see~\cite{calderbank98,gottesman97}.

In spite of the negative connotations of the term ``degenerate'', we
will argue that degeneracy is an interesting and in some sense useful
quality of a quantum code. Let us call an error nice if and only if it
acts by scalar multiplication on the stabilizer code.  Nice errors do
not require any correction, which \textit{is} a nice feature
considering the fact that operational imprecisions of a quantum
computer can introduce errors in a correction step (which is the main
reason why elaborate fault-tolerant implementations are needed).

If we assume a depolarizing channel, then errors of small weight are
more likely to occur than errors of large weight. If the stabilizer
code $Q$ is nondegenerate, then all nice errors have weight $d$ or
larger, so the most probable errors \textit{all} require (potentially
hazardous) active error correction. On the other hand, if the
stabilizer code is degenerate, then there exist nice errors of weight
less than the minimum distance. Given these observations, it would be
particularly interesting to find degenerate stabilizer codes with many
nice errors of small weight.

Although the first quantum error-correcting code by Shor was a
degenerate $[[9,1,3]]_2$ stabilizer code, it turns out that most known
quantum stabilizer code families provide pure codes. If one insists on
a large minimum distance, then nondegeneracy seems more or less
unavoidable (for example, quantum MDS codes are necessarily
nondegenerate, see~\cite{rains99}). However, the fact that most known
stabilizer codes do not have nice errors of small weight is the result
of more pragmatic considerations.

Let us illustrate this last remark with the CSS construction; similar
points can be made for other stabilizer code constructions.
Suppose we start with a classical self-orthogonal $[n,k,d]_q$ code
$C$, then one can obtain with the CSS construction an $[[n,n-2k,
\delta]]_q$ stabilizer code, where $\delta=\wt(C^\perp\setminus
C)$. Since we often do not know the weight distribution of the code
$C$, the easiest way to obtain a stabilizer code with minimum distance
at least $\delta_0$ is to choose $C$ such that its dual distance
$d^\perp\ge \delta_0$, as this ensures $\delta\ge d^\perp\ge \delta_0$. 
However, since $C\subseteq C^\perp$, the
side effect is that all nonscalar nice errors have a weight of at
least $d\ge d^\perp\ge \delta_0$.  

Our considerations above suggest a different approach. Since we would
like to have nice errors of small weight, we start with a classical
self-orthogonal code $C$ that has a small minimum distance, but is
chosen such that the vector of smallest Hamming weight in the
difference set $C^\perp\setminus C$ is large.  In general, it is of
course difficult is to find a good lower bound for the weights in this
difference set.

We illustrate this approach for degenerate quantum stabilizer codes
that are derived from classical duadic codes. Recall that the duadic
codes generalize the quadratic residue codes, see~\cite{leon84},
\cite{smid86},\cite{smid87}. We show that one can still obtain a
surprisingly large minimum distance, considering the fact we start
with classical codes that are really bad.

In Section~\ref{sec:duadic}, we recall basic properties of duadic
codes. In Section~\ref{sec:euclid}, we construct degenerate quantum
stabilizer codes using the CSS construction. Finally, in
Section~\ref{sec:hermitian}, we obtain further quantum stabilizer
codes using the Hermitian code construction.

\paragraph*{Notation} 
Throughout this paper, $n$ denotes a positive odd integer.  If $a$ is
an integer coprime to $n$, then we denote by $\ord_n(a)$ the
multiplicative order of $a$ modulo $n$. We briefly write $\qr$ to
express the fact that $q$ is a quadratic residue modulo~$n$.
We write $p^\alpha\|n$ if
and only if the integer $n$ is divisible by $p^\alpha$ but not by
$p^{\alpha+1}$.  If $\gcd(a,n)=1$, then the map $\mu_a:i\mapsto a i
\bmod n$ denotes a permutation on the set $\{
0,1,\ldots,n-1\}$. An element $c=(c_1,\ldots,c_n )\in
\F_q^n$ is said to be even-like if $\sum_i{c_i}=0$, and odd-like
otherwise. A code $C\subseteq \F_q^n$ is said to be even-like if every
codeword in $C$ is even-like, and odd-like otherwise.

\section{Classical Duadic Codes} \label{sec:duadic}
In this section, we recall the definition and basic properties of
duadic codes of length $n$ over a finite field $\F_q$ such that
$\gcd(n,q)=1$. For each choice, we will obtain a quartet of
codes: two even-like cyclic codes and two odd-like cyclic codes.

Let $S_0$, $S_1$ be the defining sets of two cyclic codes of length~$n$
over~$\F_q$ such that
\begin{compactenum}
\item  $S_0\cap S_1=\emptyset$, 
\item $S_0\cup S_1=S =
\{1,2,\ldots,n-1 \}$, and
\item $aS_i \bmod n=S_{(i+1 \bmod 2)} $ for some $a$
coprime to $n$. 
\end{compactenum}
In particular, each $S_i$ is a union of $q$-ary cyclotomic cosets
modulo $n$. Since condition 3) implies $|S_0|=|S_1|$, we have
$|S_i|=(n-1)/2$, whence $n$ must be odd. The tuple $\{ S_0,S_1,a \}$
is called a \textit{splitting} of $n$ given by the permutation~$\mu_a$. 

Let $\alpha$ be a primitive $n$-th root of unity over $\F_q$.  For
$i\in \{0,1\}$, the odd-like duadic code $D_i$ is a cyclic code of
length~$n$ over~$\F_q$  with defining set $S_i$ and generator polynomial
$$g_i(x) = \prod_{j\in S_i} (x-\alpha^j).$$ The even-like duadic code
$C_i$ is defined as the even-like subcode of $D_i$; thus, it is a
cyclic code with defining set $S_i\cup \{ 0\}$ and generator polynomial $(x-1)g_i(x)$.  We have $\dim
D_i=(n+1)/2$ and $\dim C_i=(n-1)/2$. 

\begin{theorem}\label{th:duadicexist}
Duadic codes of length $n$ over $\F_q$ exist if and only if $\qr$.
\end{theorem}
\begin{proof}
This is well-known, see, for example,~\cite[Theorem~1]{smid87}
or~\cite[Theorem~6.3.2, pages~220-221]{huffman03}.
\end{proof}

Although the weight distribution of a duadic code is not known in
general, the following well-known fact gives partial information about
the weights of odd-like codewords.

\begin{lemma}[Square Root Bound]\label{th:duadicdist}
Let $D_0$ and $D_1$ be a pair of odd-like duadic codes of length $n$
over $\F_q$.  Then their minimum odd-like weights in both codes are
same, say $d_o$.  We have
\begin{compactenum}
\item $d_o^2\geq n$,
\item $d_o^2-d_o+1\geq n$ if the splitting is given by $\mu_{-1}$. 
\end{compactenum}
\end{lemma}
\begin{proof}
See \cite[Theorem~6.5.2]{huffman03}.
\end{proof}

\section{Quantum Duadic Codes -- Euclidean Case} \label{sec:euclid}

In this section, we derive quantum stabilizer codes from classical
duadic code using the well-known CSS construction.  Recall that in the
CSS construction, the existence of an $[n,k_1]_q$ code~$C$ and an
$[n,k_2]_q$ code~$D$ such that $C\subset D$ guarantees the existence
of an $[[n,k_2-k_1,d]]_q$ quantum stabilizer code with minimum
distance $d=\min \mbox{wt}\{(D\setminus C)\cup (C^\perp\setminus
D^\perp)\}$.

\subsection{Basic Code Constructions}
Recall that two $\F_q$-linear codes $C_1$ and $C_2$ are said to be
equivalent if and only if there exists a monomial matrix $M$ and
automorphism $\gamma$ of $\F_q$ such that $C_2=C_1M\gamma$, see
\cite[page~25]{huffman03}. We denote equivalence of codes by $C_1\sim
C_2$.  For us it is relevant that equivalent codes have the same
weight distribution, see~\cite[page~25]{huffman03}.

The permutation map $\mu_a:i\mapsto ai\bmod n$ also
defines an action on polynomials in $\F_q[x]$ by $f(x)\mu_a=f(x^a)$. 
This induces an action on a cyclic code $C$ over $\F_q$ by  
$$C\mu_a = \{c(x)\mu_a \mid c(x)\in C\} =\{ c(x^a)\mid c(x)\in C\}.$$
\begin{lemma}\label{th:equivcode}
Let $C$ be a cyclic code of length $n$ over $\F_q$ with defining set $T$. If 
$\gcd(a,n)=1$, then the cyclic code $C\mu_a$ has the defining set $a^{-1}T$.
Furthermore, we have $C\mu_a\sim C$. 
\end{lemma}
\begin{proof}
This follows from the definitions, see also
\cite[Corollary~4.4.5]{huffman03} and \cite[page~141]{huffman03}.
\end{proof}

\begin{theorem}\label{th:quantumduadic1} 
Let $n$ be a positive odd integer, and let $\qr$.  There exist quantum
duadic codes with the parameters $[[n,1,d]]_q$, where $d^2\geq n$.  If
$\ord_n(q)$ is odd, then there also exist quantum duadic codes with
minimum distance $d^2-d+1\geq n$.
\end{theorem}
\begin{proof}
Let $N=\{0,1,\dots,n-1\}$.  If $\qr$, then there exist duadic codes
$C_i\subset D_i$, for $i\in \{0,1\}$. Suppose that the defining set of
$D_i$ is given by $S_i$; thus, the defining set of the even-like
subcode $C_i$ is given by $S_i\cup \{0\}$. It follows that $C_i^\perp$
has defining set $-(N\setminus (\{0 \}\cup S_{i})) = -S_{(i+1\bmod
2)}.$ Using Lemma~\ref{th:equivcode}, we obtain
$C_i^\perp=D_{(i+1\bmod 2)}\mu_{-1} \sim D_{(i+1\bmod 2)}$ and
$D_i^\perp=C_{(i+1\bmod 2)}\mu_{-1}\sim C_{(i+1\bmod 2)}$. By the CSS
construction, there exists an $[[n,(n+1)/2-(n-1)/2,d]]_q$ quantum
stabilizer code with minimum distance $d=\min\{\wt((D_i\setminus
C_i)\cup (C_i^\perp\setminus D_i^\perp)) \} $. Since $C_i^\perp\sim
D_{(i+1\bmod 2)}$ and $D_i^\perp\sim C_{(i+1\bmod 2)}$, the minimum
distance $d=\min \{\wt((D_i\setminus C_i)\cup (D_{(i+1\bmod
2)}\setminus C_{(i+1\bmod 2)}) \}$, which is nothing but the minimum
odd-like weight of the duadic codes; hence $d^2\geq n$.  If
$\ord_n(q)$ is odd, then $\mu_{-1}$ gives a splitting of $n$\cite[Lemma~5]{rushanan86}. In this
case,  Lemma~\ref{th:duadicdist} implies that the odd-like weight $d$ satisfies $ d^2-d+1 \geq
n$.~\end{proof}

In the binary case, it is possible to derive degenerate codes with
similar parameters using topological
constructions~\cite{bravyi98,freedman01,kitaev02}, but the codes do
not appear to be equivalent to the construction given here.

\subsection{Degenerate Codes} \label{sec:impure1}
The next result proves the existence of degenerate duadic quantum
stabilizer codes. This results shows that the classical duadic codes,
such as $C_i\subseteq D_i$, contain codewords of very small weight but
their set difference $D_i\setminus C_i$ (and $ C_i^\perp\setminus
D_i^\perp$) does not. First we need the following lemma, which shows
the existence of duadic codes of low distance.

\begin{lemma}\label{th:duadicevendist}
Let $n=\prod p_i^{m_i}$ be an odd integer and $\qrx{p_i}$.
If $t_i=\ord_{p_i}(q)$ and $p_i^{z_i}\| q^{t_i}-1$, and
$m_i>2z_i$, then there exists a duadic  code of length $n$ and (even-like)
minimum distance $\leq \min\{ p_i^{z_i}\} < \sqrt{n}$.
\end{lemma}
\begin{proof}
By Theorem~\ref{th:duadicexist} there exist duadic codes of lengths
$p_i^{m_i}$ and by \cite[Theorem~6]{smid87} their minimum distance, $d_i'$
is less than $p_i^{z_i}$. 
Since we know that the odd-like distance is $\geq p_i^{m_i/2} > p_i^{z_i}$, the
minimum distance must be even-like. By \cite[Theorem~4]{smid87}, there 
exists duadic codes of length
$n=\prod p_i^{m_i}$ whose minimum distance $d'\leq \min\{d_i' \} \leq \min \{p_i^{z_i} \} < \prod p_i^{m_i/2} =\sqrt{n}$. Since this is less
than the minimum odd-like distance, the minimum distance is
even-like.~\end{proof}

\begin{theorem}\label{th:impurecss}
Let $n=\prod p_i ^{m_i}$ be an odd integer and $\qrx{p_i}$.
Let $t_i=\ord_{p_{i}}(q)$, and let $z_i$ be
such that $p_i^{z_i} \| q^{t_i}-1$. Then for $m_i>2z_i$, there exists
a degenerate $[[n,1,d]]_q$ quantum code pure to $d'\leq \min
\{p_i^{z_i} \} <d$ with $d^2\geq n$. If $p_i\equiv -1\bmod 4$, then $d^2-d+1\geq n$.
\end{theorem}
\begin{proof}
From Lemma~\ref{th:duadicevendist}, we know that there exist
duadic codes of length $n$ and minimum (even-like) distance $d'\leq
\min\{p_i^{z_i}\}< \sqrt{n}$.  From Theorem~\ref{th:quantumduadic1}, we know there
exists a quantum duadic code with parameters $[[n,1,d]]$, where $d\geq \sqrt{n}>d'$.
Hence, the quantum code is degenerate.

If $p_i\equiv -1 \bmod 4$, then by
\cite[Theorem~8]{smid87}, the permutation
$\mu_{-1}$ gives a splitting for this code. Hence the odd-like distance must
satisfy $d^2-d+1$.
\end{proof}
\begin{example}
Let us consider binary quantum duadic codes of length $7^m$. Note that $2$
is a quadratic residue modulo $7$ as $4^2\equiv 2 \mod 7$. 
Since $\ord_7(2)=3$ and $7\|2^3-1$, we have $z=1$. By Theorem~\ref{th:impurecss} 
for $m\geq 2$ there exist quantum codes with the parameters $[[7^m,1,d]]_2$.
As $p=7\equiv -1 \mod 4$  we have with $d^2-d+1\geq 7^{m}$.
But, $d'$, the distance of the (even-like) duadic codes is upper bounded by $p^z=7$. 
Hence these codes are pure to $d'\leq 7$. 
Actually, using the fact that the true distance of the
even-like codes is $4$ \cite{smid87} 
we can show that the quantum codes are pure to $4$.
\end{example}

\section{Quantum Duadic Codes -- Hermitian Case}\label{sec:hermitian}
Recall that if there exists an $\F_{q^2}$-linear $[n,k,d]_{q^2}$ code $C$
such that $C^{\hdual}\subseteq C$, then there exists an $[[n,2k-n,\ge
d]]_q$ quantum stabilizer code that is pure to $d$.  In this section,
we construct duadic quantum codes using this construction.  Since 
$q^2\equiv \square\bmod n$,
duadic codes exist over $\F_{q^2}$ for all $n$, when $\gcd(n,q^2)=1$.

\subsection{Basic Code Constructions} 
\begin{lemma}\label{th:duadichdual}
Let $C_i$ and $D_i$ respectively be the even-like and odd-like duadic
codes over $\F_{q^2}$, where $i\in \{0,1\}$.  Then $C_i^\hdual = D_i$
if and only if there is a $q^2$-splitting of $n$ given by $\mu_{-q}$,
that is, $-qS_i \equiv S_{(i+1\bmod 2)}\bmod n$.
\end{lemma}
\begin{proof}
See~\cite[Theorem~4.4]{rushanan86}. 
\end{proof}

\begin{lemma}\label{th:hermitiansplitting1}
Let $n=\prod p_i^{m_i}$ be an odd integer such that $\ord_n(q)$ is odd.
Then $\mu_{-q}$ gives a splitting of $n$ over $\F_{q^2}$. In fact $\mu_{-1}$
and $\mu_{-q}$ give the same splitting. 
\end{lemma}
\begin{proof}
Suppose that $\{S_0,S_1,a\}$ be a splitting. We know that each $S_i$
is an union of some $q^2$-ary cyclotomic cosets, so $q^2S_i \equiv
S_i\bmod n$. Now $q^{\ord_n(q)}S_i \equiv S_i\bmod n$. If
$\ord_n(q)=2k+1$, then $q^{2k+1} S_i \equiv qS_i \equiv S_i\bmod n$;
hence, $\mu_q$ fixes each $S_i$ if the multiplicative order of $q$
modulo $n$ is odd.

Notice that if $\ord_n(q)$ is odd, then $\ord_n(q^2)$ is also odd. By
\cite[Lemma~5]{rushanan91}, we know that there exists a
$q^2$-splitting of $n$ given by $\mu_{-1}$ if and only if
$\ord_n(q^2)$ is odd.  Hence $-S_i \equiv S_{(i+1 \bmod 2)}\bmod
n$. Since $\mu_q$ fixes $S_i$ we have $-qS_i \equiv S_{(i+1\bmod
2)}\bmod n$; hence, $\mu_{-q}$ gives a $q^2$-splitting of $n$.

Conversely, if $\mu_{-q}$ gives a splitting of $n$, then $-qS_i\equiv
S_{(i+1\bmod 2)} \bmod n$.  But as $\mu_{q}$ fixes $S_i$ we have
$-S_i\equiv S_{(i+1\bmod 2)}\bmod n$. Therefore $\mu_{-1}$ gives the
same splitting as $\mu_{-q}$.  
\end{proof}

\begin{theorem}\label{th:hermitianduadic1}
Let $n$ be an odd integer such that $\ord_n(q)$ is odd. Then there exists
an $[[n,1,d]]_q$ quantum code with $d^2-d+1\geq n$.
\end{theorem}
\begin{proof}
By Lemma~\ref{th:hermitiansplitting1}, there exist duadic codes $C_i\subset D_i$ 
with splitting given by $\mu_{-q}$ and $\mu_{-1}$. This means that the $C_i\subseteq C_i^\hdual =D_i$ by Lemma~\ref{th:duadichdual}. Hence
there exists an $[[n,n-(n-1),d]]_q$ quantum code with
$d=\wt(D_i\setminus C_i)$. As $\mu_{-1}$ gives a splitting, we have
$d^2-d+1\geq n$ by Lemma~\ref{th:duadicdist}.~\end{proof}

\subsection{Degenerate codes}
We construct a family of degenerate quantum codes that has a large minimum
distance.
\begin{theorem}\label{th:impurehermitian}
Let $n=\prod p_i ^{m_i}$ be an odd integer with $\ord_n(q)$ odd and
every $p_i\equiv -1 \bmod 4$.
Let $t_i=\ord_{p_i}(q^2)$, and $p_i^{z_i} \| q^{2t_i}-1$. Then for $m_i>2z_i$,
there exist degenerate quantum codes with parameters $[[n,1,d]]_q$ pure to
$d'\leq \min \{p_i^{z_i} \} <d$ with $d^2-d+1\geq n$.
\end{theorem}
\begin{proof}
From Lemma~\ref{th:duadicevendist} we know that there exists an even-like duadic
code with parameters $[n,(n-1)/2,d']_{q^2}$ and $d'\leq \min \{p_i^{z_i}\}$. 

Then by \cite[Theorem~8]{smid87}, we know that for this code $\mu_{-1}$ 
gives a splitting. By Lemma~\ref{th:hermitiansplitting1}, $\mu_{-q}$ also gives a splitting for this code. 

Hence by Theorem~\ref{th:hermitianduadic1} this duadic code gives  a quantum duadic code $[[n,1,d]]_q$, which is impure as $d'\leq \min \{p_i^{z_i}\}< \sqrt{n}< d$.
\end{proof}
Finally, one can construct more quantum codes, for instance 
when $\ord_n(q)$ is even, by finding the conditions under which
$\mu_{-q}$ gives  a splitting of $n$.

\section{Conclusion}\label{sec:conclusion}
The motivation for this work was that many good quantum
error-correcting codes, such as quantum MDS codes, are typically pure
and thus require active corrective steps for all errors of small
Hamming weight. At the other extreme are decoherence free
subspaces~(see \cite{zanardi97,lidar98}) that do not require any active error
correction at all, but perform poorly in terms of minimum distance. We
pointed out that degenerate quantum codes can form a compromise,
namely they can reach larger minimum distances while allowing at least
some nice errors of low weight that do not require active error
correction.

We have constructed two families of quantum duadic codes with the
parameters $[[n,1,\geq\sqrt{n}]]_q$ and have shown that they contain
large subclasses of degenerate quantum codes. Though these codes
encode only one qubit, they are interesting because they demonstrate
that there exist families of classical codes which can give rise to
remarkable degenerate quantum codes.  Since these code are cyclic, we
know that there exist several nice errors of small weight. A more
detailed study of the weight distribution of classical duadic codes
can reveal which code are particularly interesting for quantum
error-correction.  We note that generalizations of duadic codes, such
as triadic and polyadic codes, can be used to obtain degenerate
quantum codes with higher rates.

\section*{Acknowledgment}

We thank M.H.M. Smid for sending us his thesis~\cite{smid86}.  This
research was supported by NSF CAREER award CCF~0347310, NSF grant CCR
0218582, and a Texas A\&M TITF initiative.



%

\IEEEtriggeratref{7}
\IEEEtriggercmd{\newpage}


\begin{thebibliography}{10}

\bibitem{ashikhmin01}
A.~Ashikhmin and E.~Knill.
\newblock Nonbinary quantum stabilizer codes.
\newblock {\em IEEE Trans. Inform. Theory}, 47(7):3065--3072, 2001.

\bibitem{bravyi98}
S.~B. Bravyi and A.~Y. Kitaev.
\newblock Quantum codes on a lattice with boundary.
\newblock quant-ph/9810052, 1998.

\bibitem{calderbank98}
A.R. Calderbank, E.M. Rains, P.W. Shor, and N.J.A. Sloane.
\newblock Quantum error correction via codes over {GF}(4).
\newblock {\em IEEE Trans. Inform. Theory}, 44:1369--1387, 1998.

\bibitem{freedman01}
M.H. Freedman and D.A. Meyer.
\newblock Projective plane and planar quantum codes.
\newblock {\em Found. Comput. Math.}, 1(3):325--332, 2001.

\bibitem{gottesman97}
D.~Gottesman.
\newblock Stabilizer codes and quantum error correction.
\newblock Caltech Ph.D. Thesis, eprint: quant-ph/9705052, 1997.

\bibitem{huffman03}
W.~C. Huffman and V.~Pless.
\newblock {\em Fundamentals of Error-Correcting Codes}.
\newblock University Press, Cambridge, 2003.

\bibitem{ketkar05}
A.~Ketkar, A.~Klappenecker, S.~Kumar, and P.~K. Sarvepalli.
\newblock Nonbinary stabilizer codes over finite fields.
\newblock quant-ph/0508070, 2005.

\bibitem{kitaev02}
A.~Kitaev.
\newblock Topological quantum codes and anyons.
\newblock In {\em Quantum computation: a grand mathematical challenge for the
  twenty-first century and the millennium (Washington, DC, 2000)}, volume~58 of
  {\em Proc. Sympos. Appl. Math.}, pages 267--272. Amer. Math. Soc.,
  Providence, RI, 2002.

\bibitem{leon84}
J.~Leon, J.~Masley, and V.~Pless.
\newblock Duadic codes.
\newblock {\em IEEE Trans. Inform. Theory}, 30(5):709--714, 1984.

\bibitem{lidar98}
D.A. Lidar, I.L. Chuang, and K.B. Whaley.
\newblock Decoherence-free subspaces for quantum-computation.
\newblock {\em Phys. Rev. Letters}, 81:2594--2597, 1998.

\bibitem{rains99}
E.M. Rains.
\newblock Nonbinary quantum codes.
\newblock {\em IEEE Trans. Inform. Theory}, 45:1827--1832, 1999.

\bibitem{rushanan86}
J.J. Rushanan.
\newblock {\em Topics in Integral Matrices and Abelian Group Codes}.
\newblock Ph.{D}. thesis, California Institute of Technology, 1986.

\bibitem{rushanan91}
J.J. Rushanan.
\newblock Duadic codes and difference sets.
\newblock {\em J.\ Combin.\ Theory Ser.\ A}, 57:254--261, 1991.

\bibitem{smid86}
M.~H.~M. Smid.
\newblock {\em On Duadic Codes}.
\newblock Dept. Math., Tech. Univ., Netherlands, 1986.

\bibitem{smid87}
M.~H.~M. Smid.
\newblock Duadic codes.
\newblock {\em IEEE Trans. Inform. Theory}, 3:432--433, 1987.

\bibitem{zanardi97}
P.~Zanardi and M.~Rasetti.
\newblock Noiseless quantum codes.
\newblock {\em Phys. Rev. Lett.}, 79:3306, 1997.

\end{thebibliography}

\end{document}